# Uncontrolled source titles: When Nature, Science, and PNAS meet Scopus

Weishu Liu wsliu08@163.com
https://orcid.org/0000-0001-8780-6709
School of Management, Zhejiang University of Finance and Economics, Hangzhou 310018, Zhejiang, China

Ruifeng Zhang zhangruifeng@zufe.edu.cn
School of Information Technology and Artificial Intelligence, Zhejiang University of Finance and Economics, Hangzhou 310018, Zhejiang, China

Haifeng Wang wanghf@shisu.edu.cn corresponding author
https://orcid.org/0000-0003-4967-7971
School of Business and Management, Shanghai International Studies University, Shanghai, China

**Acknowledgments**
We utilized ChatGPT to refine the language of this manuscript, and subsequently, we conducted a comprehensive manual review and revision. We are responsible for any errors.

**Declaration of conflicting interest**
The authors declare that there is no conflict of interest.

**Ethical approval and informed consent statements**
Not applicable.

**Data availability statement**
Due to copyright restrictions, we are not permitted to publicly share the data. Readers can obtain detailed data through their subscription to Scopus and Web of Science Core Collection.

**Abstract:** Bibliographic databases such as Scopus and Web of Science play a critical role in literature discovery, systematic reviews, bibliometric analyses, and research evaluation, yet their reliability depends on effective control of the quality of the bibliographic data. This study examines the problem of uncontrolled source titles in Scopus by comparing five possibly adopted journal-level retrieval strategies across three flagship journals—Nature, Science, and PNAS. Substantial discrepancies across retrieval results indicate persistent fragmentation in the source title field. A comparative analysis using the Web of Science Core Collection shows that no comparable degree of uncontrolled source-title variation is observed for the same journals. The study further explores possible mechanisms, including inheritance of different title forms from (different) upstream providers, automated parsing and OCR-related artifacts, and insufficient cross-validation. Based on these findings, the study proposes recommendations at both the user and database-provider levels to improve source-title normalization and strengthen source-level metadata governance, with implications for the accuracy and reproducibility of journal-level retrieval and evaluative uses.

## 1. Introduction

Bibliographic databases such as Scopus and Web of Science constitute core infrastructures of contemporary scholarly communication [1-2]. They support literature discovery, enable large-scale systematic reviews and meta-analyses, and underpin bibliometric analyses, while increasingly serving as critical supporting tools for research assessment, funding allocation, and evidence-based science policymaking [3-6]. Across these diverse applications, reliable journal-level identification is foundational. For example, many systematic reviews and meta-analyses depend on accurate source title filtering, whereas evaluative and policy-oriented analyses rely on aggregating publications and citations at the journal level. These practices implicitly assume that source-level metadata (journal titles and their associated identifiers, e.g., ISSN, CODEN, and internal source IDs) are consistently controlled, complete, and unambiguous within bibliographic databases.

Although Scopus and Web of Science are often self-proclaimed as high-quality databases [7-8], a substantial body of literature has documented the quality problems in article-level metadata in these two platforms [9-10]. Widely reported issues include errors and incompleteness in author names [11-13], institutional affiliations [14-16], abstracts [17-18], funding acknowledgments [19-21], references [22], document types [23-24], and DOIs [25-26]. These problems are generally understood as structural challenges arising from the scale, heterogeneity, and automated processing of bibliographic data [19, 27].

In contrast, journal-level metadata describes comparatively stable publication entities that are typically associated with fixed titles and persistent identifiers such as ISSN, and are therefore often assumed to be easier to standardize and govern. In recent years, cross-database comparative studies have begun to doubt this assumption by revealing inconsistencies in journal-level metadata across bibliographic databases, including incompleteness of the ISSN information and variations in journal titles [28-32]. These studies have demonstrated that the same journal may be represented differently in different systems, complicating data integration and comparative analysis. However, this line of research has primarily focused on discrepancies between databases, implicitly assuming that journal-level metadata are internally consistent within each database. Whether this assumption holds in practice (after excluding source title changes) has received little systematic examination, despite its importance for retrieval reliability and analytical reproducibility.

This relative neglect is consequential. Within a curated database such as Scopus, uncontrolled source titles (primarily manifested as persistent variation in source-title assignment) may affect large numbers of records while remaining largely invisible to users. Unlike over-retrieval, which typically produces obvious false positives, under-retrieval caused by fragmented or inconsistent source titles is difficult to detect. Users may therefore unknowingly work with incomplete datasets while assuming comprehensive coverage. Moreover, because source titles simultaneously function as search keys, filtering criteria, and aggregation units, inconsistencies at the source level can compromise multiple journal-level retrieval strategies, even when stable identifiers are nominally available.

This study addresses this gap by empirically examining uncontrolled source titles within Scopus. Focusing on three well-established multidisciplinary journals—Nature, Science, and Proceedings of the National Academy of Sciences of the United States of America (PNAS)—we compare five possibly adopted journal-level retrieval strategies: source-title searching, source-title filtering, ISSN-based retrieval, Scopus Source ID matching, and CODEN-based retrieval. These three journals are selected because they have long publication histories, stable and well-defined titles, and are widely recognized in scholarly communication, which reduces ambiguity in journal identity. These three journals serve as conservative test cases: if source-title fragmentation occurs here—where title stability is well established—it suggests that similar problems may also occur in journals with less standardized or more ambiguous titles. Accordingly, discrepancies observed across retrieval strategies may reflect inconsistencies in source-title normalization within Scopus, while ambiguity arising from journal boundaries or title changes is minimized by the selection of these three journals.

We hypothesize that the source-title inconsistencies observed in Scopus may reflect broader limitations in source-level metadata governance rather than isolated indexing anomalies, although the extent to which similar problems occur in other journals remains to be investigated. By demonstrating how persistent source-title inconsistency fragments journal-level retrieval even for highly standardized journals, this study highlights an underappreciated source of error in bibliographic searching and journal-based analysis. The findings contribute to ongoing discussions on metadata quality, database transparency, and the reliability of bibliographic infrastructures in literature retrieval, evidence synthesis, and evaluative and policy-oriented uses.

## 2. Data and methods

This study employs a comparative case study design to examine uncontrolled source titles in Scopus by comparing the results of multiple journal-level retrieval strategies applied to the same journals. These strategies do not create inconsistencies but differentially expose existing source-level fragmentation. Differences across retrieval results are interpreted as evidence of fragmentation or inconsistency in source-level metadata. Three elite multidisciplinary journals—Nature, Science, and Proceedings of the National Academy of Sciences of the United States of America (PNAS)—are selected as conservative cases. As long-established and highly standardized journals, they are less likely to suffer from uncontrolled source titles.

Five possibly used journal-level retrieval strategies in Scopus are implemented: exact source-title search, source-title filtering, ISSN-based retrieval, Scopus Source ID matching, and CODEN-based retrieval (Table 1). For comparison, parallel retrievals are conducted in the Web of Science Core Collection. The used Web of Science Core Collection includes Science Citation Index Expanded from 1900, Social Sciences Citation Index from 1900, Arts & Humanities Citation Index from 1975, Conference Proceedings Citation Index from 1990 (both Science and Social Sciences & Humanities versions), and Book Citation Index from 2005 (both Science and Social Sciences & Humanities versions). To reduce the influence of ongoing database updates, the analysis is restricted to publications published prior to 2025. Searches were conducted via the advanced search interface of these two platforms on February 3rd, 2026, without limits on document type, language, or subject area. The results obtained on this date were used for the main analyses reported in Tables 2–7. On September 20th, 2026, we conducted additional verification searches through the Scopus web interface and API. For retrieval operations that functioned as intended, the differences in record counts were small relative to the February 2026 totals. These results were used for verification only and did not replace the February 2026 results used in the main analyses. However, during the September verification, the EXACTSRCTITLE field (Method #1) in the Scopus web interface did not consistently follow its documented matching behavior, although it had worked as expected in the earlier searches. Conversely, the Scopus API did not correctly execute the source-title filtering operation (Method #2).

Table 1 Five possible search strategies in Scopus

| Method No. | Search field | Explanation |
|---|---|---|
| #1 | Exact Source Title (field tag: EXACTSRCTITLE) | Search the source title in the "Exact Source Title" field. Exact source title searches sources that contain the exact words but not variations of the search terms. |
| #2 | Source title filter (field tag: EXACTSRCTITLE) | Use the source title filter. |
| #3 | ISSN (field tag: ISSN) | Search the ISSN and EISSN in the "ISSN" field. |
| #4 | Source Identifier (field tag: SRCID ) | Search the Source ID created by Scopus in the "Source Identifier" field. |
| #5 | CODEN (field tag: CODEN ) | Search the CODEN in the "CODEN" field. |

## 3. Discrepancies of retrieval strategies and uncontrolled source titles in Scopus

This section examines how uncontrolled source titles manifest in Scopus through three empirical cases: Nature, Science, and PNAS. For each journal, retrieval results obtained via five strategies are compared, and observed discrepancies are analyzed in terms of record counts and assigned source-title variants.

### 3.1. The case of Nature

Applying the term “Nature” in the exact source title field in Scopus retrieves nearly 640,000 records (Method #1; Table 2). This figure far exceeds the actual publication output of Nature, reflecting large-scale over-retrieval caused by the partial match algorithm adopted by Scopus: any indexed source title containing the word “nature” is retrieved. As a result, exact source-title searching in Scopus cannot be considered a valid strategy for journal-level retrieval in this case. Applying the Scopus source title filter substantially reduces the result set to 333,636 records. Comparable counts are obtained via ISSN-based retrieval with 333,631 records and Scopus Source ID (namely 21206) matching with 333,636 records. At first glance, these near-identical figures suggest internal consistency across identifier-based approaches. However, closer inspection reveals meaningful discrepancies.

Table 2 Retrieval results for Nature

| Method No. | Search query | Number of records |
|---|---|---|
| #1 | PUBYEAR < 2025 AND EXACTSRCTITLE ( "Nature" ) | 639745 |
| #2 | PUBYEAR < 2025 AND ( LIMIT-TO ( EXACTSRCTITLE , "Nature" ) ) | 333636 |
| #3 | PUBYEAR < 2025 AND ISSN ( 00280836 OR 14764687 ) | 333631 |
| #4 | PUBYEAR < 2025 AND SRCID ( 21206 ) | 333636 |
| #5 | PUBYEAR < 2025 AND CODEN(NATUA) | 69199 |

ISSN-based retrieval yields five fewer records than the source-title-filtered and Source ID-based strategies. Manual inspection shows that four of these records are published by Nature in 2005, for which ISSN metadata are missing in the bibliographic information provided in Scopus, while the fifth is a book titled "Nature" published by Routledge. Notably, Scopus assigns both the journal and this book to the same internal Source ID (namely 21206), suggesting a failure to distinguish between fundamentally different publication entities. All four Nature records lacking ISSN metadata in Scopus are MEDLINE-derived. While this pattern may suggest a connection between record provenance and identifier incompleteness, the small number of affected records precludes any definitive conclusion.

The CODEN-based strategy exhibits the most severe coverage limitation. Only 69,199 records are retrieved—approximately one-fifth of the journal's total output indexed in Scopus. CODEN information is almost entirely absent for records published prior to 1996, and remains incomplete thereafter: approximately 14.5% of post-1995 Nature records still lack CODEN values. These findings indicate that CODEN cannot be regarded as a reliable identifier for comprehensive or longitudinal journal-level retrieval in Scopus. Although Chemical Abstracts Service continues to assign CODEN upon request (https://www.cas.org/services/coden), the identifier has fallen into substantial disuse in modern publishing practice. CODEN-based retrieval was included because it persists in legacy workflows and older bibliographic records.

### 3.2. The case of Science

Applying the same retrieval strategies to the journal Science yields even greater divergence across methods (Table 3). An exact source-title search for "Science" retrieves more than 5.3 million records, reflecting extreme over-retrieval due to the generic nature of the term. As in the case of Nature, this strategy is unsuitable for identifying records published in Science. When the Scopus source title filter is applied, 188,321 records are retrieved. In contrast, ISSN-based and Source ID-based strategies yield substantially larger result sets (191,545 and 191,552 records, respectively), exceeding the filtered result by more than 3,000 records. As observed for Nature, the CODEN field also exhibits limited completeness for Science, particularly for records published prior to 1996, further constraining its suitability for comprehensive retrieval.

Table 3 Retrieval results for Science

| Method No. | Search query | Number of records |
|---|---|---|
| #1 | PUBYEAR < 2025 AND EXACTSRCTITLE( "Science" ) | 5384674 |
| #2 | PUBYEAR < 2025 AND ( LIMIT-TO ( EXACTSRCTITLE , "Science" ) ) | 188321 |
| #3 | PUBYEAR < 2025 AND ISSN ( 00368075 OR 10959203 ) | 191545 |
| #4 | PUBYEAR < 2025 AND SRCID ( 23571 ) | 191552 |
| #5 | PUBYEAR < 2025 AND CODEN(SCIEA) | 70353 |

To investigate the sources of these discrepancies, records retrieved by Methods #2–#5 were further decomposed by the source titles assigned in Scopus. As demonstrated in Table 4, six distinct title variants associated with Science can be identified. In addition to the standard title "Science", records appear under other variants or even incorrect forms can be identified via ISSN-based and Source ID-based strategies: "Science New York N Y" with 2951 records, "Science Science" with 232 records, "Sciences" with 35 records, "Sciecne" with 14 records and "Sciece" with 9 records. In total, 3241 records retrieved via ISSN or source ID do not carry the typical source title, directly explaining the reduced recall of the source-title-filtered strategy. In addition, among all records retrieved via the CODEN field, two lack source-title information (namely "undefined" in Scopus). Given their non-standard document title strings and early publication years (1969 and 1970), the underlying causes of the missing source titles cannot be determined.

Table 4 Decomposition by source titles (Science)

| Source title in Scopus | Use the source title filter | Search by ISSN | Search by source ID | Search by CODEN |
|---|---|---|---|---|
| Science | 188321 | 188304 | 188311 | 70351 |
| Science New York N Y | 0 | 2951 | 2951 | 0 |
| Science Science | 0 | 232 | 232 | 0 |
| Sciences | 0 | 35 | 35 | 0 |
| Sciecne | 0 | 14 | 14 | 0 |
| Sciece | 0 | 9 | 9 | 0 |
| Undefined | 0 | 0 | 0 | 2 |
| Total | 188321 | 191545 | 191552 | 70353 |

Further inconsistencies are observed even among records assigned the standard title, as demonstrated in the second row of Table 4. Compared to ISSN-based retrieval, the Source ID strategy identifies seven additional records. Six of them lack ISSN metadata in the bibliographic information provided in Scopus, while the seventh is a book titled "Science" published by Acumen Publishing. As in the case of Nature, Scopus assigns both the journal Science and a same-titled book to a single Source ID (namely 23571), again conflating distinct publication entities. Conversely, the source title filter retrieves ten records not captured by the Source ID strategy. These ten records are all MEDLINE-derived published during 1980s and originate from a discontinued journal with the same title but a different ISSN (i.e., 0193-4511).

### 3.3. The case of PNAS

For PNAS, inconsistencies are less pronounced but remain clearly observable (Table 5). Exact source-title searching and source-title filtering both retrieve 151,486 records, whereas ISSN- and Source ID-based strategies retrieve slightly larger sets (151,998 and 152,005 records, respectively). CODEN-based retrieval again shows reduced coverage, capturing 142,275 records and omitting approximately 6.4% of PNAS records indexed in Scopus.

Table 5 Retrieval results for PNAS

| Method No. | Search query | Number of records |
|---|---|---|
| #1 | PUBYEAR < 2025 AND EXACTSRCTITLE ( "Proceedings of the National Academy of Sciences of the United States of America" ) | 151486 |
| #2 | PUBYEAR < 2025 AND ( LIMIT-TO ( EXACTSRCTITLE , "Proceedings of the National Academy of Sciences of the United States of America" ) ) | 151486 |
| #3 | PUBYEAR < 2025 AND ISSN ( 00278424 OR 10916490 ) | 151998 |
| #4 | PUBYEAR < 2025 AND SRCID ( 21121 ) | 152005 |
| #5 | PUBYEAR < 2025 AND CODEN(PNASA) | 142275 |

Analysis of assigned source titles of the used strategies reveals five distinct variants associated with PNAS (Table 6). Most notably, 512 records are indexed under a truncated title (namely, “Proceedings of the National Academy of Sciences of the United States of”) for the retrieved results of the ISSN- and Source ID-based strategies. Three other abbreviated or simplified forms also appear for the Source ID-based strategy ("Proceedings National Academy of Sciences Usa" with five records, "Proc Nat ACAD of SCI U S A" with one record, and "Proceedings National Academy of Science U S A" with one record). As in the previous cases, the presence of these variants reduces the recall of source-title-based filtering, while ISSN- and Source ID-based strategies continue to retrieve the affected records.

Table 6 Decomposition by source titles (PNAS)

| Source title in Scopus | Use the source title filter | Search by ISSN | Search by source ID | Search by CODEN |
|---|---|---|---|---|
| Proceedings of the National Academy of Sciences of the United States of America | 151486 | 151486 | 151486 | 142275 |
| Proceedings of the National Academy of Sciences of the United States of | 0 | 512 | 512 | 0 |
| Proceedings National Academy of Sciences Usa | 0 | 0 | 5 | 0 |
| Proc Nat ACAD of SCI U S A | 0 | 0 | 1 | 0 |
| Proceedings National Academy of Science U S A | 0 | 0 | 1 | 0 |
| Total | 151486 | 151998 | 152005 | 142275 |

Although the magnitude of inconsistency for PNAS is smaller than that observed for Science, its presence is nevertheless significant. Even for a journal with a formally standardized and distinctive title, Scopus exhibits persistent variation in source-title assignment.

## 4. A comparative check on Web of Science Core Collection

Web of Science Core Collection (WoSCC) is another widely used multidisciplinary bibliographic database and is therefore examined here for comparative purposes. In contrast to Scopus, only three journal-level retrieval strategies are available in Web of Science due to interface constraints: (1) searching the publication title field, (2) applying the publication title filter, and (3) ISSN-based retrieval. Table 7 summarizes the number of records retrieved for Nature, Science, and PNAS using these three approaches.

Table 7 Retrieval results in Web of Science Core Collection

| Journal | Method No. | Search field | Search query | Number of records |
|---|---|---|---|---|
| Nature | #1 | Publication Title (field tag: SO) | FPY=1900-2024 AND SO= "Nature" | 269292 |
| | #2 | Publication Title filter (field tag: SO) | FPY=1900-2024 and then set the publication title filter as "Nature" | 269292 |
| | #3 | ISSN (field tag: IS) | FPY=1900-2024 AND IS= ( 00280836 OR 14764687 ) | 269165 |
| Science | #1 | Publication Title (field tag: SO) | FPY=1900-2024 AND SO= "Science" | 168353 |
| | #2 | Publication Title filter (field tag: SO) | FPY=1900-2024 and then set the publication title filter as "Science" | 168353 |
| | #3 | ISSN (field tag: IS) | FPY=1900-2024 AND IS= ( 00368075 OR 10959203 ) | 168353 |
| PNAS | #1 | Publication Title (field tag: SO) | FPY=1900-2024 AND SO= "Proceedings of the National Academy of Sciences of the United States of America" | 153454 |
| | #2 | Publication Titles filter (field tag: SO) | FPY=1900-2024 and then set the publication title filter as "Proceedings of the National Academy of Sciences of the United States of America" | 153454 |
| | #3 | ISSN (field tag: IS) | FPY=1900-2024 AND IS= ( 00278424 OR 10916490 ) | 161885 |

### 4.1. The case of Nature

For Nature, searching the publication title field retrieves 269,292 records for the period 1900–2024. An identical number of records is retrieved when using the publication title filter. This consistency indicates that Web of Science applies a full-match algorithm to the publication title field, rather than the partial match used in Scopus. As a result, the publication title field in Web of Science does not suffer from the large-scale over-retrieval observed in Scopus when querying titles with generic terms. ISSN-based retrieval for Nature yields 269,165 records, omitting 127 records retrieved by the title-based approaches. Manual inspection shows that these omitted records are associated with the book series "Cahiers de la Revue de théologie et de philosophie". This series includes a source titled "Nature" published in 1996, which carries a different ISSN (namely 0250-6971).

### 4.2. The case of Science

For the journal Science, all three retrieval strategies yield an identical number of records (168,353). We did not find any anomalies in the publication title field for this journal. All records with ISSNs are collected correctly in Web of Science.

### 4.3. The case of PNAS

For PNAS, searching the full publication title retrieves 153,454 records, a number identical to that obtained via the publication title filter. ISSN-based retrieval, however, yields 8,431 additional records. These records correspond to issues published between 1980 and 1984, which Web of Science indexes under two distinct titles: Proceedings of the National Academy of Sciences of the United States of America–Biological Sciences (8,022 records) and Proceedings of the National Academy of Sciences of the United States of America–Physical Sciences (409 records). According to the National Library of Medicine, PNAS issues during this period were published in two parts with consecutive page numbering. In this case, the discrepancy arises not from uncontrolled source-title variation but from historically differentiated journal sections that are explicitly documented and systematically indexed. However, from a user perspective, title-based retrieval for PNAS in WoSCC risks omitting a substantial number of records indexed under historically differentiated journal sections, while ISSN-based retrieval, though more complete—also requires disambiguation. In this sense, the practical retrieval challenge is not entirely absent in WoSCC.

### 4.4. A summary

Taken together, the Web of Science results contrast sharply with those observed for Scopus. First, exact title searching in Web of Science does not generate large-scale false positives, due to the use of a full-match algorithm in the publication title field. Second, source-title variants such as truncated forms, typographical errors, or duplicated titles were not observed to a comparable extent in the three WoSCC cases examined. Third, cases where ISSN-based retrieval yields additional records can be traced to historically differentiated journal sections, rather than to uncontrolled variation within a single source identity. Nevertheless, these findings do not imply that WoSCC is free from metadata-related retrieval limitations, as illustrated by the PNAS case. Finally, the substantial differences in absolute record counts between Scopus and Web of Science for the three journals (Tables 2, 3, and 5 versus Table 7) are likely attributable to multiple factors, including differences in coverage period (e.g., approximately 27,000 pre-1900 records for Nature and 12,000 for Science are retrievable in Scopus but not in WoSCC), differences in document type indexing policy, and possible metadata errors, duplicate records, or omissive indexing [33-35]. As these differences may mask underlying data quality issues, a systematic investigation falls outside the scope of the current study but warrants future research.

## 5. Possible mechanisms underlying uncontrolled source titles in Scopus

The analyses above document persistent variation in source-title assignment within Scopus for three prominent multidisciplinary journals. While the empirical results establish the existence, scale, and concrete manifestations of uncontrolled source titles, they do not by themselves, identify possible causes. In the absence of direct access to Scopus's internal indexing workflows and normalization rules, this section discusses several plausible mechanisms that could account for the observed patterns. These mechanisms are inferred from the empirical characteristics of the affected records, their metadata provenance, and documented practices in large-scale bibliographic database construction, and should therefore be interpreted as possible explanatory hypotheses rather than confirmed causal explanations.

### 5.1. Inheritance of different title forms from (different) upstream providers

Given the observed variability across journal-level retrieval strategies, one possible mechanism underlying uncontrolled source titles in Scopus is the inheritance of inconsistent bibliographic metadata from multiple upstream sources. Unlike Web of Science, which developed through decades of continuous journal indexing, Scopus was launched in 2004 and rapidly established broad coverage by incorporating large volumes of retrospective records from heterogeneous external sources, including thousands of publishers and some existing databases [35-37]. As a result, records originating from providers such as MEDLINE, Embase, and Compendex may reflect divergent metadata standards and title normalization practices [35, 37]. Under these conditions, identical journals may be introduced into Scopus under different source-title strings depending on record provenance.

The empirical analyses show that part of records associated with anomalous source titles are derived from MEDLINE. For example, an examination of the top 20 highly cited records among the 2,951 records indexed under the variant title "Science New York N Y", spanning publication years from 1945 to 2024, shows that all of them are labeled as sourced from MEDLINE. According to the NLM Catalog, "Science (New York, N.Y.)" represents the uniform title for Science within MEDLINE. The Scopus variant closely matches the MEDLINE uniform title, suggesting that upstream title variants may not be fully reconciled during data ingestion.

In the absence of comprehensive normalization procedures that enforce a single controlled source title per journal across all ingestion pathways, such inherited variants may persist and accumulate over time, potentially contributing to fragmented source identities. While direct verification of this mechanism is not possible, the MEDLINE provenance labels observed in the examined records are consistent with this possible explanatory hypothesis.

### 5.2. Automated parsing and OCR-related artifacts

In addition to provider-related inconsistencies, several of the source-title variants identified in the preceding analyses are consistent with artifacts introduced during automated metadata extraction from unstructured or scanned records. For retrospectively indexed records and legacy content, structured publisher-supplied metadata are not always available, requiring bibliographic information to be extracted from scanned documents or unstructured textual sources. Under such conditions, journal titles may be inferred from headers, footers, cover pages, or citation strings, increasing the likelihood of truncation, duplication, or character-level misspellings.

The results based on sampled records for Science and PNAS are consistent with this interpretation. In the case of Science, several rare source-title variants (e.g., character-level misspellings) appear only in small numbers and may be consistent with OCR-related or other metadata-processing errors. For PNAS, 512 records are indexed under a truncated title—"Proceedings of the National Academy of Sciences of the United States of"—with publication years concentrated in 1949–1951 and 1961–1965. Manual inspection of the top 20 highly cited records shows that all these records are labeled as sourced from MEDLINE. At the same time, other MEDLINE-derived records published during the same periods are assigned the correct source title. This coexistence suggests that records from identical journals and time spans may have been processed under different extraction conditions, ingestion batches, or parsing workflows.

### 5.3. Insufficient cross-validation across metadata fields

Beyond string-level variation, the fragmentation patterns documented above also suggest limitations in cross-validation across source-level metadata fields, particularly where publication type and identifiers are not jointly enforced. In a well-controlled bibliographic system, source titles would ideally be validated jointly against publication type, publisher information, and stable identifiers such as ISSN, thereby ensuring consistent assignment of records to publication entities and preventing the conflation of distinct sources.

The empirical analyses reveal several patterns consistent with incomplete cross-field validation. Records sharing identical ISSN values are sometimes associated with multiple source-title variants, while journals and monographs sharing the same or similar titles are occasionally assigned the same Scopus Source ID despite clear differences in publication type and publisher. These patterns suggest that title-based similarity may, in some instances, outweigh other metadata fields during source assignment.

Without effective cross-validation across metadata fields, two complementary problems may emerge: fragmentation, in which records from a single journal are distributed across multiple source identities, and conflation, in which distinct publication entities are merged under a single source identity. Although the internal decision rules governing source assignment in Scopus are not publicly documented, the observed coexistence of fragmentation and conflation across multiple high-profile journals indicates that limitations in cross-field validation represent a plausible explanation for the persistence of uncontrolled source titles.

Taken together, these factors suggest that uncontrolled source titles in Scopus are not merely the byproduct of scale or historical depth but may reflect a broader governance issue at the source-metadata level, at least insofar as the three journals examined here allow such an inference to be drawn. The problem may therefore lie not only in the initial introduction of variation, which is arguably unavoidable in large, heterogeneous databases, but also in the incomplete validation, reconciliation, and retroactive correction of source titles over time.

## 6. Potential impacts of uncontrolled source titles in Scopus

Uncontrolled source titles in Scopus, primarily manifested as persistent source-title inconsistency, have implications that extend beyond isolated retrieval errors. The presence of these inconsistencies in three conservative test cases suggests that similar problems may also affect journals with less standardized titles, greater linguistic diversity, or more complex publication histories. Because source titles operate as key access points for searching, filtering, aggregation, and evaluation, inconsistencies at this level can propagate across multiple downstream uses of bibliographic data.

First, uncontrolled source titles reduce recall in journal-level retrieval. When records from the same journal are distributed across multiple title variants, users relying on source-title filters or exact title matching may unknowingly retrieve incomplete datasets. For example, relying on the source-title filter may fail to retrieve over 3,200 records in Science. Unlike over-retrieval, which often produces visible false positives, under-retrieval caused by fragmented source titles is largely invisible and therefore difficult for users to detect or correct.

Second, source-title inconsistency can distort bibliometric analyses and derived indicators. Fragmentation of records across multiple title variants may lead to underestimation of publication counts, citation totals, and other indicators calculated from incomplete journal-level retrieval results. For example, filtering by the source title "Science" retrieves 1,490 records published in Science in 2024; however, an additional 448 records indexed under the variant source title "Science New York N Y" are missed by this approach, representing a 23% undercount (448/1938). Such distortions can propagate to institutional or disciplinary analyses that employ journal-based inclusion criteria, resulting in indicators that reflect metadata artifacts rather than underlying scholarly activity.

Third, uncontrolled source titles undermine the reproducibility of bibliometric research. Studies nominally examining the same journal may produce different results depending on the retrieval strategies employed and the extent to which source-title variants are harmonized. Without explicit documentation and reconciliation of these variants, analytical outcomes become difficult to replicate and compare.

Finally, because bibliographic databases such as Scopus are frequently used in research evaluation and decision-making, source-title inconsistency has broader implications for evaluative practices and user trust. Journal-based assessments may be affected in opaque ways, and persistent inconsistencies (even for prominent journals) challenge expectations regarding the stability and transparency of curated bibliographic infrastructures.

## 7. Discussion

The central finding of this study, based on three conservative test cases, is that source-title fragmentation in Scopus can occur even under favorable conditions—journals with long histories, distinctive titles, and global recognition. If a journal as prominent and standardized as Science can accumulate records under six distinct title variants, including outright misspellings, the problem is unlikely to be confined to obscure or ambiguously named journals, though how far it extends across the broader journal population remains to be examined systematically. The three conservative cases collectively suggest several possible explanations for source-title fragmentation: metadata inherited from (different) upstream providers without adequate normalization, character-level errors introduced during automated parsing or OCR processing of historical records, and the lack of effective mechanisms to validate source titles against stable identifiers such as ISSN and to retroactively correct affected records. Prior studies have documented similar data quality issues for the field of author names [38-39], affiliations [40], countries/territories [15] and even publication dates [41] in Scopus or Web of Science. However, source-title metadata—despite functioning as the primary access point for journal-level retrieval, aggregation, and evaluation — has received comparatively little attention in this literature.

These findings connect to two broader themes: authority control and identifier governance. In library and information science, authority control refers to the practice of establishing a single authorized form for an entity and linking all variant forms to it — a mechanism designed precisely to prevent the kind of fragmentation documented here. Yet large-scale bibliographic databases that rely on automated ingestion from heterogeneous upstream providers appear prone to the emergence of "distributed source identities," where the same journal is represented as multiple distinct metadata entities, thereby undermining the collocating function that bibliographic systems are designed to serve. This vulnerability is not a technical inevitability. Scopus already assigns a stable Source ID to each indexed journal, which in principle provides the identifier infrastructure needed to normalize source title strings; that variant titles persist nonetheless — even within records sharing the same Source ID — suggests that existing mechanisms may not fully enforce consistency between identifier-based source entities and the title strings attached to individual records. Other bibliographic databases, such as OpenAlex [42–43], provide useful opportunities for future comparisons of source-level metadata governance and normalization practices. The fragmentation documented here may therefore reflect limitations in source-level metadata governance and is, in principle, potentially addressable.

## 8. Recommendations for users and database providers

The findings of this study, while based on three conservative test cases, suggest that the problems associated with uncontrolled source titles in Scopus are unlikely to be entirely inherent or unavoidable, and may be mitigated through improvements at both the user and database levels.

From the perspective of users, exclusive reliance on a single journal-level retrieval strategy is often insufficient. Exact source-title searches or source-title filters alone may lead to incomplete datasets when source-title variants are present. Accordingly, combining multiple retrieval strategies, particularly source-title filtering in conjunction with ISSN- and Source ID-based queries—can improve recall. Users are further advised to inspect retrieved records for variant source titles, including truncated, abbreviated, or misspelled forms, and to harmonize these variants into a standardized journal title prior to analysis. Clear and transparent documentation of retrieval strategies and any harmonization steps is essential for ensuring reproducibility, particularly for historical records and records associated with external metadata sources, such as the MEDLINE-derived examples examined here.

From the perspective of Scopus and other database providers, improvements in source-title normalization and identifier completeness would likely yield substantial benefits. Systematic reconciliation of variant source titles, supported by authoritative title registries and stable identifiers, could substantially reduce fragmentation at the source level. Although Scopus may already have undertaken efforts in this direction, a large number of inconsistencies can still be identified. Targeted retroactive correction of missing ISSN and CODEN metadata particularly for legacy records would strengthen identifier-based retrieval and aggregation. In addition, explicit and consistently enforced differentiation between publication entities sharing the same title, such as journals and monographs or discontinued serials, would prevent conceptual conflation that currently undermines precision. More generally, systematic cross-validation across metadata fields (e.g., source title, publication type, publisher, and identifiers) would enhance the coherence and reliability of source-level metadata.

## 9. Limitations and future research

This study has several limitations that also suggest directions for future research. First, the analysis focuses on only three high-profile journals—Nature, Science, and PNAS—which, while intentionally chosen as conservative test cases, limit the generalizability of the findings to other journals with different disciplinary, linguistic, or historical characteristics. The extent of similar problems in other journals remains to be investigated. Second, the possible provenance analysis is restricted to the top 20 highly cited records among affected records, constituting a non-random and potentially biased sample that limits the evidential weight of the possible provenance hypothesis; a more systematic provenance analysis, where additional information is available, would be a valuable next step. Third, the analysis is based on observable retrieval outcomes and metadata patterns, without access to Scopus's internal indexing workflows. Moreover, complete historical provenance and processing records may not be available even within Scopus. The proposed mechanisms should be interpreted as plausible hypotheses rather than confirmed causes. As bibliographic databases are continuously updated, exact record counts may change over time, although the detailed retrieval procedures reported here enable independent repetition of the searches. Future studies could extend this approach to a broader range of journals and bibliographic systems to further assess the scope and drivers of uncontrolled source titles.